%% file: mybib.tex
\documentclass[a4paper]{article}
\usepackage{microtype}
\usepackage{url}
\usepackage{INTERSPEECH2022}
\usepackage{amsmath}
\usepackage{siunitx}

\input{math_commands.tex}

\usepackage{color}
\usepackage[colorlinks,linkcolor=black,urlcolor=black,citecolor=black]{hyperref}

\title{VoiceFixer: A Unified Framework for High-Fidelity Speech Restoration}
\name{
Haohe Liu$^{1,*,\dag\thanks{* The first two authors contributed equally to this work.}\thanks{\dag~Part of this work was done during the internship at ByteDance}}$,
Xubo Liu$^{1,*}$, 
Qiuqiang Kong$^2$, 
Qiao Tian$^2$, 
Yan Zhao$^2$, \\ 
DeLiang Wang$^3$
Chuanzeng Huang$^2$, 
Yuxuan Wang$^2$
}

\address{
  $^1$Centre for Vision, Speech and Signal Processing~(CVSSP), University of Surrey, UK\\
  $^2$Speech, Audio, and Music Intelligence~(SAMI) Group, ByteDance, China\\
  $^3$Department of Computer Science and Engineering, The Ohio State University, USA
  }
\email{hl01486@surrey.ac.uk, kongqiuqiang@bytedance.com, dwang@cse.ohio-state.edu}

\begin{document}

\maketitle
\begin{abstract}
    Speech restoration aims to remove distortions in speech signals. Prior methods mainly focus on a single type of distortion, such as speech denoising or dereverberation. However, speech signals can be degraded by several different distortions simultaneously in the real world. It is thus important to extend speech restoration models to deal with multiple distortions. In this paper, we introduce VoiceFixer, a unified framework for high-fidelity speech restoration. VoiceFixer restores speech from multiple distortions~(e.g., noise, reverberation, and clipping) and can expand degraded speech~(e.g., noisy speech) with a low bandwidth to \num{44.1} kHz full-bandwidth high-fidelity speech. We design VoiceFixer based on~(1)~an analysis stage that predicts intermediate-level features from the degraded speech, and~(2)~a synthesis stage that generates waveform using a neural vocoder. Both objective and subjective evaluations show that VoiceFixer is effective on 
   severely degraded speech, such as real-world historical speech recordings. Samples of VoiceFixer are available at \url{https://haoheliu.github.io/voicefixer}.

\end{abstract}

\section{Introduction}
Human speech often suffers from distortions such as background noise, room reverberations, or clipping from low-quality devices. Those distortions degrade the perceptual quality of human listeners. Speech restoration is a task to restore degraded speech to high-quality speech, which is useful in a wide range of applications such as online  meeting~\cite{enhancement-online-meeting-defossez2020real} and hearing aids~\cite{enhancement-hearning-aids-van2009speech}.

Previous speech restoration methods mainly focus on a single type of distortion, such as speech denoising~\cite{loizou-speech-enhancement-2007speech}, dereverberation~\cite{zhang2021weighted}, super-resolution~\cite{audio-supre-resolution-SR-kuleshov2017audio}, and declipping~\cite{declipping-overview-zavivska2020survey}. However, in the real world, speech signals can be degraded by several different distortions simultaneously. These mismatches limit the performance of these systems. Several works have explored restoring speech with multiple distortions, such as noise and reverberation~\cite{ai2021denoising, su2020hifi}. But other distortions such as low-resolution and clipping receive less attention, despite their significant impacts on speech perceptual quality.

Speech fidelity is important to perceptual quality. However, existing methods show limited performance on high-fidelity speech restoration. For example, for a noisy speech with low bandwidth, although the speech denoising method could remove noises, the restored speech would be still in low fidelity. One way to address this issue is to concatenate speech restoration methods~(e.g., denoising) with the speech super-resolution method. However, this approach has limitations such as increasing computational cost and accumulating the artifacts introduced by each speech restoration model. To our knowledge, restoring low-bandwidth speech with multiple distortions has not been studied in the literature. 

This paper introduces VoiceFixer, a unified framework for high-fidelity speech restoration. VoiceFixer restores speech from multiple distortions~(e.g., noise, reverberation, and clipping) and could expand distorted speech with a low bandwidth between 1 kHz and 22.05 kHz to a full-bandwidth high-fidelity speech signal. We design VoiceFixer based on a two-stage strategy:~(1) an analysis stage that performs mel spectrogram estimation;~(2) a synthesis stage that generates the speech signal from the estimated mel spectrogram. Compared to the conventional speech restoration methods that operate on spectrogram or waveform, VoiceFixer uses the low dimensional mel spectrogram as the intermediate-level feature, which alleviates the difficulties of restoring multiple distortions simultaneously. In addition, neural vocoders~\cite{hifi-gan-kong2020hifi} are usually trained on large-scale speech datasets. This provides prior knowledge on synthesizing waveform from low-dimensional mel spectrogram. The contributions of this paper are listed as follows: 

\begin{list}{\labelitemi}{\leftmargin=1em}
\item We present VoiceFixer, a unified framework for \num{44.1} kHz high-fidelity speech restoration. VoiceFixer can restore degraded speech from multiple distortions~(e.g., noise, reverberation, clipping, and low-bandwidth).
\item Evaluation result shows the effectiveness of VoiceFixer, which achieves a \num{0.256} higher mean opinion scores~(MOS) than the baseline method. 
\item We release the pre-trained model and source code\footnote{\url{https://github.com/haoheliu/voicefixer\_main}} of VoiceFixer to encourage future research.
\end{list}

\begin{figure*}[t] 
    \centering
    \vspace{-0.6em} 
    \includegraphics[page=11,width=1.\linewidth]{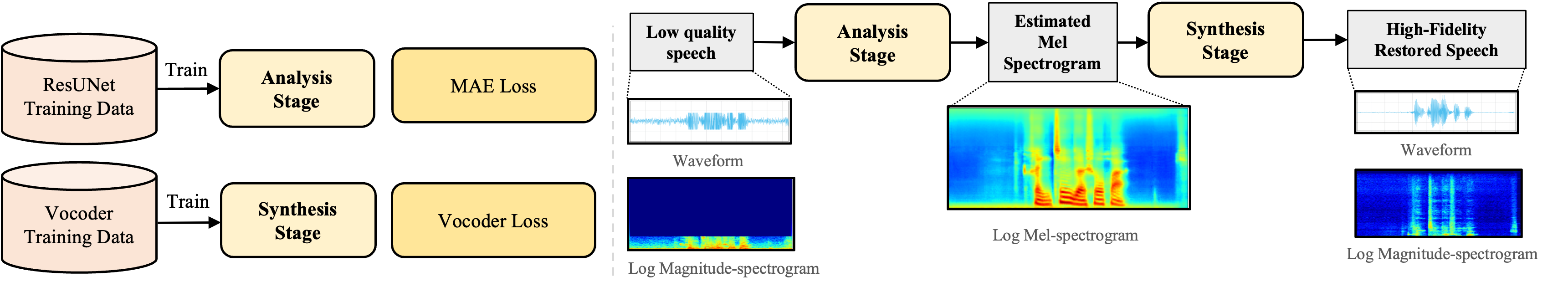}
    \caption{Overview of the proposed VoiceFixer framework. The analysis module and synthesis module are trained separately.}
    \label{fig-overview-VoiceFixer}
\end{figure*}

The rest of this paper is organized as follows. \Secref{section-problem-formulations} introduces the formulations of speech distortions we addressed. \Secref{sec:approach} describes the architecture of our proposed VoiceFixer. Experiments are presented in \Secref{section-experiments}. In \Secref{section-conclusion}, we summarize this study and discuss our future directions.

\section{Problem Formulation}
\label{section-problem-formulations}
We denote a segment of a speech signal as $s \in \R^{L}$, where $ L $ is the number of samples in the segment. We model the speech distortion process as function $d(\cdot)$. The degraded speech $x \in \R^{L}$ thus can be written as $x = d(s)$. Speech restoration aims to restore high-quality speech $ \hat{s} $ from $ x $ by $\hat{s} = f(x)$, where $ f(\cdot) $ is the restoration function and can be viewed as an inverse approximation of $ d(\cdot) $. The target of the restoration function is to estimate $s$ by restoring $\hat{s}$ from the degraded speech $x$.

\label{section-distortion-simulation}

Distortion modeling is an important step to simulate training data when building speech restoration systems. Previous works model distortions in a sequential order~\cite{two-stage-tan2020audio,joint-aec-noise-suppression-shu2021joint}. Similarly, we model the distortion $ d(\cdot) $ as a composite function:

\begin{equation}
    \label{eq-distortion-composition}
    d(x)=d_1\circ d_2\circ ... d_Q(x), d_q \in \sD, q=1,2,...,Q, 
\end{equation}
where $ \circ $ stands for function composition and $ Q $ is the number of distortions consisted in $ d(\cdot) $. $ \sD = \{d_v(\cdot)\}_{v=1}^{V} $ is the set of distortion types, where $ V $ is the total number of types. \Eqref{eq-distortion-composition} describes the procedure of compounding different distortions from $ \sD $ in a sequential order. The four types of speech distortions we addressed in this work are introduced as follows.



\noindent
\textbf{Additive noise} is one of the most common distortion and can be modeled by the addition between speech $ s $ and noise $n \in \R^{L}$:
\begin{equation}
    \label{eq-noisy}
    d_{\text{noise}}(s) = s + n.
\end{equation}



\noindent
\textbf{Reverberation} is caused by the reflections of signal within a space. Reverberation makes speech signals sound distant and blurred. It can be modeled by convolving speech signals with a room impulse response filter~(RIR) $r$:
\begin{equation}
    \label{eq-reverberation}
    d_{\text{rev}}(s)=s * r,
\end{equation}

\noindent where $ * $ stands for convolution operation.

\noindent
\textbf{Clipping} distortion refers to the clipped amplitude of audio signals when their amplitude exceeds the maximum level.
Clipping can be modeled by restricting signal amplitudes within a range $[-\reta,+\reta]$:
\begin{equation}
    \label{eq-clipping}
    d_{\text{clip}}(s)=\mathrm{max}(\mathrm{min}(s,\reta),-\reta),\reta\in[0,1].
\end{equation}
In the frequency domain, the clipping effect produces harmonic components in the high-frequency part and degrades speech intelligibility accordingly.

\noindent
\textbf{Low-bandwidth} distortion refers to the limited bandwidth in the audio recordings caused by low sampling rate or defects in the recording device. We follow the description in~\cite{heming-towards-sr-wang2021towards} to produce low-bandwidth distortions but add more filter types~\cite{filter-generalization-overfitting-sulun2020filter}. After designing a low pass filter $h$, we first convolve it with $ \vs $ to avoid the aliasing phenomenon. Then we perform resampling on the filtered result from the original bandwidth $o$ to a lower bandwidth $u$:
\begin{equation}
    \label{eq-low-resolution}
    d_{\text{low\_bw}}(s)=\mathrm{Resample}(s*h, o, u).
\end{equation}

\section{Approach}
\label{sec:approach}
The two-stage strategy of VoiceFixer is formulated as follows: 
\begin{equation}
    \label{eq-voice-fixer-two-stages1}
    f: x \mapsto z,
\end{equation}
\begin{equation}
    \label{eq-voice-fixer-two-stages2}
    g: z \mapsto \hat{s}.
\end{equation}
\Eqref{eq-voice-fixer-two-stages1} denotes the analysis stage of VoiceFixer where a distorted speech $ x $ is mapped into a intermediate-level feature $ z $. \Eqref{eq-voice-fixer-two-stages2} denotes the synthesis stage of VoiceFixer, which synthesizes $ z $ to the restored speech $ \hat{s} $. The overview of VoiceFixer framework is depicted in Figure \ref{fig-overview-VoiceFixer}.

\subsection{Analysis stage} 

The goal of the analysis stage is to predict the intermediate representation $ \vz $, which can be used later to recover the speech signal.
In our study, we choose the mel spectrogram as the intermediate representation. Mel spectrogram has been widely used in tasks such as speech enhancement \cite{maiti2020speaker} and audio synthesis \cite{mel-gan-kumar2019melgan, liu2021conditional}. The frequency dimension of the mel spectrogram is usually much smaller than the magnitude spectrogram calculated using short-time-fourier-transform~(STFT), thus working on mel-scale can reduce the dimension of feature space and offer a more tractable restoration process. The objective of the analysis stage is to restore the mel spectrogram of the target signals, which can be written as follows:
\begin{equation}
    \label{eq-voice-fixer-restoration-mel}
    \hat{S}_{\text{mel}} = f_{\text{mel}}(X_{\text{mel}}; \alpha) \odot~(X_{\text{mel}} + \eps),
\end{equation}
\noindent where $ X_{\text{mel}} $ is the mel spectrogram of $ x $. It is calculated by $ X_{\text{mel}} = \left| X \right| W $, where $|X|$ is the magnitude spectrogram of $x$ and $ W $ is a set of mel filter banks. The columns of $ W $ are not divided by the numbers of mel bands, because this will make the restoration model difficult to recover the high-frequency part. The mapping function $ f_{\text{mel}}(\cdot ; \alpha) $ is the mel-restoration mask-estimation model parameterized by $ \alpha $. $ X_{\text{mel}} $ is added with a minimum value $\eps$ before multipling with the output of $ f_{\text{mel} }$. $\eps$ is set to \num{1e-8} in this work to avoid zero values in $ X_{\text{mel}}$.

We use ResUNet~\cite{kong2021decoupling, liu2021cws} to model the analysis stage. ResUNet consists of six encoder and six decoder blocks. There are skip connections between encoder and decoder blocks at the same level. Both encoder and decoder blocks have a similar structure of four residual convolutions. Each residual convolutional consists of a batch normalization, a leakyReLU activation, and a two-dimensional convolutional operation. We utilize average pooling and transpose convolution for the upsampling and downsampling in the encoder and decoder blocks. We will refer to ResUNet as UNet in the remaining parts. We optimize the model in the analysis stage using the MAE loss between the estimated and the target mel spectrogram, $ \hat{S}_{\text{mel}} $ and $ S_{\text{mel}} $:

\begin{equation}
    \mathcal{L}_{\text{MAE}} = \left\| \hat{S}_{\text{mel}} - S_{\text{mel}} \right\|_1.
\end{equation}

\subsection{Synthesis stage} 

We realize the synthesis stage with a neural vocoder, which synthesizes the mel spectrogram into waveform, as denoted in the following \Eqref{eq-voice-fixer-restoration}:
\begin{equation}
    \label{eq-voice-fixer-restoration}
    \hat{s} = g(X_{\text{mel}}; \beta),
\end{equation}
where $g(\cdot; \beta)$ stands for the vocoder model parameterized by $\beta$. 
The number of speakers used for the training of vocoder is much larger than that used in the analysis stage, which increases the robustness of VoiceFixer when generalizing to unseen speakers. We employ a pre-trained\footnote{\url{https://github.com/haoheliu/voicefixer}} time and frequency domain-based generative adversarial network~(TFGAN)~\cite{tian2020tfgan} as a vocoder. TFGAN achieves strong performance on 44.1 kHz speaker-independent speech vocoding, which will be discussed in detail in \Secref{section-experiments}. 

\section{Experiments}
\label{section-experiments}
We conduct two types of experiments to evaluate the performance of VoiceFixer:~(1) High-fidelity speech restoration from simultaneously appearing noise, reverberation, clipping, and low-bandwidth distortions;~(2) Single type restoration from speech with only one type of distortion~(e.g., denoising). In the following sections, we first describe the experimental data preparation, then present the results of these two experiments. The test sets used in this section are publicly available\footnote{\url{https://zenodo.org/record/5528144}}.

\subsection{Experimental data preparation}
\label{sec:Experimental-data}
Training a speech restoration system relies on pairs of distorted speech and clean speech. In the high-fidelity speech restoration task, we simulate speech with multiple distortions. As introduced in Section \ref{section-problem-formulations}, we simulate four types of speech distortion: additive noise, reverberation, clipping, and low-bandwidth. Three types of datasets are used for the simulation, including clean speech, noise data, and room impulse response~(RIR). 
Note that clipping and low-resolution distortion only need the clean speech dataset for simulation and do not depend on other datasets. We introduce the three types of datasets as follows.

\noindent
\textbf{Clean speech} we used is based on VCTK~\cite{vctk-yamagishi2019cstr}, which is a multi-speaker English corpus that consists of \num{110} speakers with different accents. The version of VCTK we used is \num{0.92}. Following the setups in other studies~\cite{nu-wave-lee2021nu}, speakers p280 and p315 are omitted for the technical issues. The remaining part is split into a training set VCTK-Train with \num{98} speakers and a testing set VCTK-Test with the last \num{8} speakers. The remaining \num{2} speakers are omitted as they appear in the test set for denoising.

\noindent
\textbf{Noise data} we used is based on two datasets. The first one is VCTK-Demand~(VD)~\cite{vctk-demand-valentini2017noisy}. VD contains a training part VD-Train and a testing part VD-Test. Both parts contain clean speech and noisy speech data. To obtain the noise data from VD, we minus each noisy data in VD-Train with its corresponding clean part to get the noise dataset VD-Noise for training. 
The second noise dataset we use is the TUT urban acoustic scenes 2018 dataset~\cite{dcase-dataset-mesaros2018multi}, which contains \num{89} hours of high-quality recording from \num{10} acoustic scenes~(e.g., airport). This dataset contains a development part and an evaluation part. We only use the evaluation part~(DCASE-Eval) for the simulation of the test set for high-fidelity speech restoration.

\noindent
\textbf{Room impulse response} is randomly simulated to add reverberation effect on 44.1 kHz speech. Simulation is performed using an open-source tool\footnote{\url{https://github.com/sunits/rir\_simulator\_python}}. All the related parameters are randomized, including the size of the room, the placement of the microphone and the sound source, the RT60 value, and the pickup pattern of the microphone. In total, 43239 filters are simulated, in which we randomly split out 5000 filters as the test set RIR-Test and named the other 38239 filters as RIR-Train.

\subsection{High-Fidelity speech restoration}

\subsubsection{Data sets and distortion modeling}
\label{sec-dataset-hifires}
In this task, we simulate low-quality speech with four distortions in the training and test set, including noise, reverberation, clipping, and low bandwidth. Training data is simulated on the fly based on the speech data in VCTK-Train, the noise in VD-Noise, and RIR in RIR-Train. We set up the parameters of each distortion to be completely random to better cover the real-world cases. The test set we used in high-fidelity speech restoration, HiFi-Res, is constructed based on the clean speech in VCTK-Test, the noise in DCASE-Eval, and RIR in RIR-Test. HiFi-Res consists of \num{501} three seconds utterances with similar random distortions simulated as the training process. 
We first generate the distortions following specific order: reverberation, noise, and clipping. Then the degraded speech is low-pass filtered and down-sampled to an arbitrary low sampling rate between \num{2} kHz to \num{44.1} kHz. Details of the distortion modeling in this work are made available on GitHub\footnote{\url{https://github.com/haoheliu/voicefixer\_main}}.

\subsubsection{Experiment details}
All the audio files in our datasets are resampled to \num{44.1} kHz sampling rate. We calculate STFT using the Hanning window with a window length of \num{2048} and a hop size of \num{441}. The mel filterbank we used consists of \num{128} filters. For training, We use Adam optimizer with $\beta_1=0.5,\beta_2=0.999$, an initial learning rate of \num{3e-4} and a batch size of \num{24}. The first \num{1000} steps are warmup steps, during which the learning rate grows linearly from \num{0} to \num{3e-4}. The learning rate is scheduled for decay by \num{0.9} every \num{400} hours of training data. We trained our model using four Nvidia-V100-32GB GPUs for two days. 


\subsubsection{Baseline systems}
We mainly use four baseline systems in the experiment. We implemented an UNet-based system~(Baseline-UNet) for the high-fidelity speech restoration task, which structure is similar to the analysis module of VoiceFixer. It performs restoration by estimating STFT of the high-quality speech and reusing the phase of the degraded speech, which is a common approach in previous speech restoration systems~\cite{choi2020phase}. As for the Oracle-Mel system, we directly use the target mel spectrogram as input to the vocoder to simulate the case when the analysis module works ideally. So, Oracle-Mel marks the theoretical upper bound of the VoiceFixer performance. For the Target system, scores are calculated using the ground truth clean speech. Conversely, the Unprocessed system evaluated directly on the distorted speech. 

\subsubsection{Evaluation metrics}
We use both objective and subjective evaluation metrics. The objective metrics including log-spectral distance~(LSD)~\cite{lsd-erell1990estimation}, wideband perceptual evaluation of speech quality~(PESQ-wb)~\cite{pesq-rix2001perceptual}, and structural similarity~(SSIM)~\cite{ssim-wang2004image}. Since neural vocoders generate waveforms directly from mel spectrograms, even with the same perceptual quality, the generated waveforms may not align with the target waveform in the time domain. This mis-alignment can considerably degrade the objective metrics, as is often the case in generative model~\cite{nu-gan-kumar2020nu}. Nevertheless, we report the model performance on these objective metrics for reference.


We use mean opinion scores~(MOS) as the subjective evaluation metric and invite eight internal language experts in ByteDance to perform evaluation. Their task is to rate the overall speech quality of an audio clip with a score between \num{1}~(bad) to \num{5}~(excellent). Each system has \num{38} samples for evaluation. We average the MOS values across all language experts as the final result.


\subsubsection{Evaluation results}
\begin{figure}[t] 
  \centering
  \includegraphics[width=0.90\linewidth]{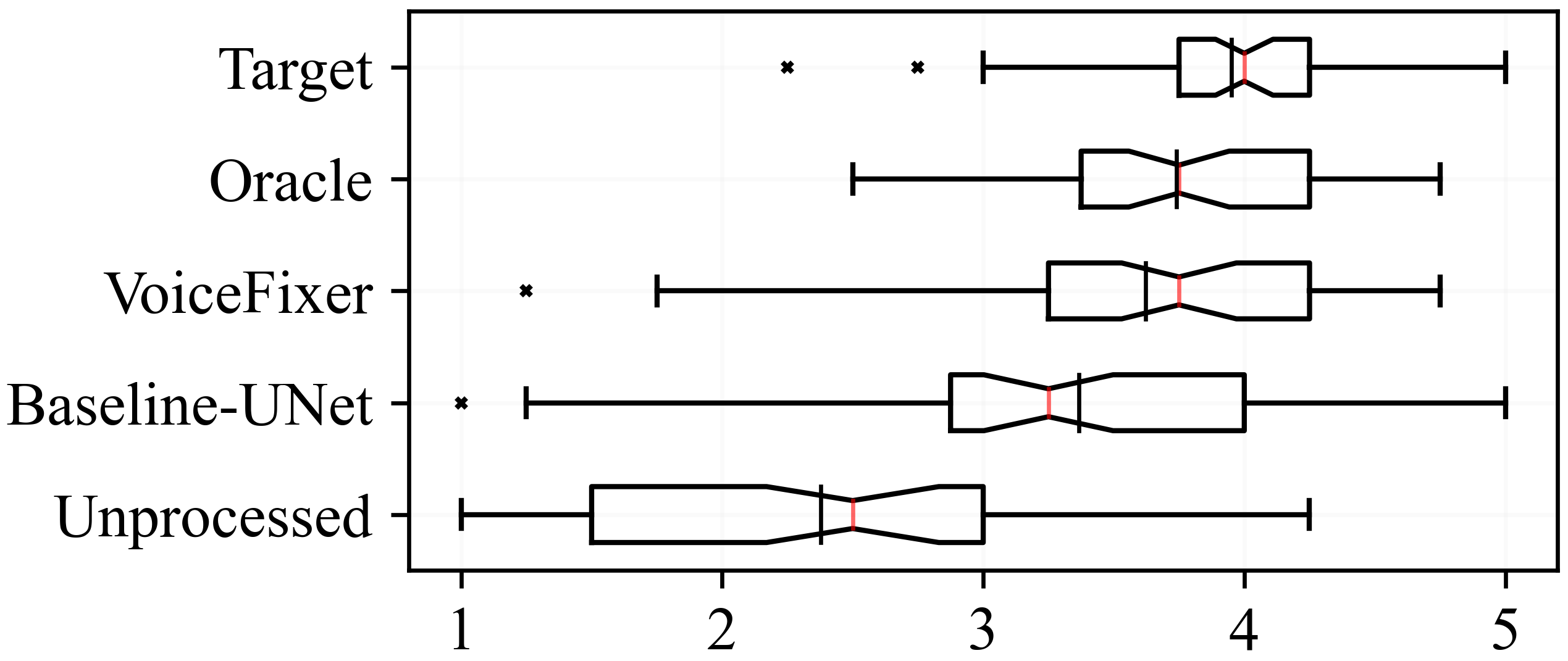}
  \caption{Box plot of the MOS scores on HiFi-Res test set. Red and black vertical lines represent median and mean values.}
  \label{fig-2}
\end{figure}

Table \ref{table-1} shows the experimental results and Figure \ref{fig-2} depicts the box plot of the MOS scores. The Oracle-Mel system achieves a MOS score of \num{3.74}, which is close to the Target MOS of \num{3.95}, indicating that the vocoder performs well in the synthesis stage. We observe that VoiceFixer obtains \num{0.256} higher MOS score than that of Baseline-UNet and is only \num{0.11} lower than the Oracle-Mel, demonstrating its good performance for high-fidelity speech restoration. Although VoiceFixer performs worse on PESQ-wb and SSIM metrics, it has a much better MOS score than Baseline-UNet. This result shows that the improvement in subjective metrics in VoiceFixer is not always consistent with objective evaluations.

\begin{table}[t]
\centering
\caption{Evaluation result on the high-fidelity speech restoration test set HiFi-Res. Higher PESQ-wb, SSIM, MOS value indicates better performance, while LSD is the opposite. The best value for each metric is shown in bold. }
\begin{tabular}[\linewidth]{c c c c c} 
 \hline
 Models & PESQ-wb & LSD & SSIM & MOS\\ 
 \hline
 Unprocessed & 1.94 & 2.00 & 0.64 & 2.38\\ 
 Oracle-Mel & 2.52 & 0.91 & 0.74 & 3.74\\ 
 Target & 4.64 & 0.01 & 1.00 & 3.95\\ \hline
 Baseline-UNet & \textbf{2.67} & 1.01 & \textbf{0.79} & 3.37\\ 
 VoiceFixer & 2.05 & \textbf{1.01} & 0.71 & \textbf{3.62}\\ 
 \hline
\end{tabular}
\label{table-1} 

\end{table}

\subsection{Single type restoration}
To further demonstrate the effectiveness of VoiceFixer, we conduct two benchmark speech restoration experiments: speech denoising and speech declipping. 

\subsubsection{Denoising}
\label{sec:denoising}
For speech denoising, we  
evaluate the model performance on VD-Test~(as described in \Secref{sec:Experimental-data}). VD-Test contains 824 utterances from a female speaker and a male speaker. The test set is simulated at four SNR levels, which are \num{17.5} dB, \num{12.5} dB, \num{7.5} dB, and \num{2.5} dB. The original data is sampled at \num{48} kHz. We downsample it to \num{44.1} kHz to fit our experiments. We adopt three recent methods  SEGAN~\cite{segan-pascual2017segan}
, WaveUNet~\cite{waveunet-macartney2018improved}, and the model trained with weakly labeled data ~\cite{enhancement-qiuqiang-weakly-labeled-kong2021speech}~(referred to WL-Model) as baseline methods.

We use the objective metric PESQ-wb~\cite{pesq-rix2001perceptual} and the subjective metric MOS to evaluate model performance. Experimental results are shown in Table \ref{table-2}. The PESQ-wb score of VoiceFixer reaches \num{2.43}, higher than SEGAN, WaveUNet, and WL-Model. The MOS evaluations demonstrate that VoiceFixer outperforms the baseline speech denoising model Baseline-UNet. In addition, we observe that VoiceFixer even outperforms Oracle-Mel and achieves the same level as Target on the MOS scores. This is because the restored results of the VoiceFixer contain more energy in the high-frequency part, which potentially leads to a better perceptual quality for the listener. 

\subsubsection{Declipping}
For the declipping task, we compare VoiceFixer with a state-of-the-art synthesis-based method SSPADE~\cite{sspade-zavivska2019proper}. To evaluate the model performance, we create a test set DECLI based on VCTK-Test~(as described in \Secref{sec:Experimental-data}). DECLI is constructed by first normalizing the amplitude of VCTK-Test into $[-1,1]$, and then simulating clipping on each audio with two clipping levels \num{0.25} and \num{0.1}. This resulted in two declipping test sets, each containing \num{2937} clipped and clean speech audios.

We adopt MOS as the subjective metric and STOI~\cite{taal2011algorithm} as the objective metric. A higher STOI value indicates better performance. Experimental results are shown in Table \ref{table-3}. VoiceFixer outperforms SSPADE on MOS by \num{0.04} and \num{1.25} in \num{0.25} and \num{0.1} clipping levels, respectively. The higher performance on MOS demonstrates a better perceptual quality restoration offered by VoiceFixer on speech declipping.  

\begin{table}[t]
\centering
\caption{Evaluation result on the VD-Test test set. Superscript $*$ indicates the model is only trained on a single restoration task.}
\begin{tabular}[\linewidth]{c c c} 
 \hline
 Models & PESQ-wb & MOS\\ 
 \hline
 Unprocessed & 1.97 & 3.20\\ 
 Oracle-Mel & 2.85 & 3.64\\ 
 Target & 4.50 & 3.69\\ \hline
 $^{*}$SEGAN \cite{segan-pascual2017segan} & 2.16 & /\\
 $^{*}$WaveUNet \cite{waveunet-macartney2018improved} & 2.40 & /\\
 $^{*}$WL-Model \cite{enhancement-qiuqiang-weakly-labeled-kong2021speech} & 2.28 & /\\
 Baseline-UNet & \textbf{2.82} & 3.64\\ 
 VoiceFixer & 2.43 & \textbf{3.69}\\ 
 \hline
\end{tabular}
\label{table-2} 
\end{table}

\begin{table}[]
\centering
\caption{Evaluation result on the declipping test set DECLI.}
\begin{tabular}{@{}c|cc|ccll}
\hline
Clipping Level & \multicolumn{2}{c|}{0.25} & \multicolumn{2}{c}{0.1}\\ \hline
Models         & STOI        & MOS        & STOI        & MOS\\ \hline
Unprocessed    & 0.95        & 2.56       & 0.89        & 2.72\\ 
Oracle-Mel     & 0.81        & 3.44       & 0.81        & 3.42\\ 
Target         & 1.00           & 3.42       & 1.00           & 3.49\\  \hline
$^{*}$SSPADE \cite{sspade-zavivska2019proper}         & \textbf{0.98}        & 3.34       & 0.92        & 2.63\\ 
Baseline-UNet  & 0.97        & 3.38       & \textbf{0.96}        & 3.38\\ 
VoiceFixer     & 0.82        & \textbf{3.38}       & 0.80        & \textbf{3.38}\\ 
\hline
\end{tabular}
\label{table-3} 
\end{table}

\section{Conclusion}
\label{section-conclusion}
In this study, we propose VoiceFixer, an effective approach for high-fidelity speech restoration. VoiceFixer consists of an analysis stage modeled by a ResUNet and a synthesis stage using a TFGAN. The two stages can also be replaced by other deep learning models. The subjective evaluation results show that VoiceFixer achieves superior performance on high-fidelity speech restoration from distortions such as noise, reverberation, clipping, and low bandwidth. In the future, VoiceFixer will be extended to more types of distortions.
\clearpage


\bibliographystyle{IEEEtran}

\bibliography{mybib}

\end{document}

%% file: math_commands.tex
\usepackage{amsmath,amsfonts,bm}

\def\Secref#1{Section~\ref{#1}}

\def\eqref#1{equation~\ref{#1}}
\def\Eqref#1{Equation~\ref{#1}}

\def\1{\bm{1}}

\def\eps{{\epsilon}}

\def\reta{{\textnormal{$\eta$}}}

\def\vs{{\bm{s}}}

\def\vz{{\bm{z}}}

\DeclareMathAlphabet{\mathsfit}{\encodingdefault}{\sfdefault}{m}{sl}
\SetMathAlphabet{\mathsfit}{bold}{\encodingdefault}{\sfdefault}{bx}{n}

\def\sD{{\mathbb{D}}}

\newcommand{\R}{\mathbb{R}}



%% file: mybib.bbl
\begin{thebibliography}{10}
\providecommand{\url}[1]{#1}
\csname url@samestyle\endcsname
\providecommand{\newblock}{\relax}
\providecommand{\bibinfo}[2]{#2}
\providecommand{\BIBentrySTDinterwordspacing}{\spaceskip=0pt\relax}
\providecommand{\BIBentryALTinterwordstretchfactor}{4}
\providecommand{\BIBentryALTinterwordspacing}{\spaceskip=\fontdimen2\font plus
\BIBentryALTinterwordstretchfactor\fontdimen3\font minus
  \fontdimen4\font\relax}
\providecommand{\BIBforeignlanguage}[2]{{%
\expandafter\ifx\csname l@#1\endcsname\relax
\typeout{** WARNING: IEEEtran.bst: No hyphenation pattern has been}%
\typeout{** loaded for the language `#1'. Using the pattern for}%
\typeout{** the default language instead.}%
\else
\language=\csname l@#1\endcsname
\fi
#2}}
\providecommand{\BIBdecl}{\relax}
\BIBdecl

\bibitem{enhancement-online-meeting-defossez2020real}
A.~Defossez, G.~Synnaeve, and Y.~Adi, ``Real time speech enhancement in the
  waveform domain,'' \emph{arXiv:2006.12847}, 2020.

\bibitem{enhancement-hearning-aids-van2009speech}
T.~Van~den Bogaert, S.~Doclo, J.~Wouters, and M.~Moonen, ``Speech enhancement
  with multichannel wiener filter techniques in multimicrophone binaural
  hearing aids,'' \emph{The Journal of the Acoustical Society of America}, vol.
  125, no.~1, pp. 360--371, 2009.

\bibitem{loizou-speech-enhancement-2007speech}
P.~C. Loizou, \emph{Speech enhancement: theory and practice}.\hskip 1em plus
  0.5em minus 0.4em\relax CRC press, 2007.

\bibitem{zhang2021weighted}
J.~Zhang, M.~D. Plumbley, and W.~Wang, ``Weighted magnitude-phase loss for
  speech dereverberation,'' in \emph{Proceedings of the IEEE Conference on
  Acoustics, Speech, and Signal Processing}, 2021, pp. 5794--5798.

\bibitem{audio-supre-resolution-SR-kuleshov2017audio}
V.~Kuleshov, S.~Z. Enam, and S.~Ermon, ``Audio super resolution using neural
  networks,'' \emph{arXiv:1708.00853}, 2017.

\bibitem{declipping-overview-zavivska2020survey}
P.~Z{\'a}vi{\v{s}}ka, P.~Rajmic, A.~Ozerov, and L.~Rencker, ``A survey and an
  extensive evaluation of popular audio declipping methods,'' \emph{IEEE
  Journal of Selected Topics in Signal Processing}, vol.~15, no.~1, pp. 5--24,
  2020.

\bibitem{ai2021denoising}
Y.~Ai, H.~Li, X.~Wang, J.~Yamagishi, and Z.~Ling,
  ``Denoising-and-dereverberation hierarchical neural vocoder for robust
  waveform generation,'' in \emph{2021 IEEE Spoken Language Technology Workshop
  (SLT)}.\hskip 1em plus 0.5em minus 0.4em\relax IEEE, 2021, pp. 477--484.

\bibitem{su2020hifi}
J.~Su, Z.~Jin, and A.~Finkelstein, ``{HiFi-GAN}: High-fidelity denoising and
  dereverberation based on speech deep features in adversarial networks,''
  \emph{arXiv:2006.05694}, 2020.

\bibitem{hifi-gan-kong2020hifi}
J.~Kong, J.~Kim, and J.~Bae, ``{HiFi-GAN}: Generative adversarial networks for
  efficient and high fidelity speech synthesis,'' \emph{arXiv:2010.05646},
  2020.

\bibitem{two-stage-tan2020audio}
K.~Tan, Y.~Xu, S.-X. Zhang, M.~Yu, and D.~Yu, ``Audio-visual speech separation
  and dereverberation with a two-stage multimodal network,'' \emph{IEEE Journal
  of Selected Topics in Signal Processing}, vol.~14, no.~3, pp. 542--553, 2020.

\bibitem{joint-aec-noise-suppression-shu2021joint}
X.~Shu, Y.~Zhu, Y.~Chen, L.~Chen, H.~Liu, C.~Huang, and Y.~Wang, ``Joint echo
  cancellation and noise suppression based on cascaded magnitude and complex
  mask estimation,'' \emph{arXiv:2107.09298}, 2021.

\bibitem{heming-towards-sr-wang2021towards}
H.~Wang and D.~Wang, ``Towards robust speech super-resolution,'' \emph{IEEE/ACM
  Transactions on Audio, Speech, and Language Processing}, vol.~29, pp.
  2058--2066, 2021.

\bibitem{filter-generalization-overfitting-sulun2020filter}
S.~Sulun and M.~E. Davies, ``On filter generalization for music bandwidth
  extension using deep neural networks,'' \emph{IEEE Journal of Selected Topics
  in Signal Processing}, vol.~15, no.~1, pp. 132--142, 2020.

\bibitem{maiti2020speaker}
S.~Maiti and M.~I. Mandel, ``Speaker independence of neural vocoders and their
  effect on parametric resynthesis speech enhancement,'' in \emph{Proceedings
  of the IEEE Conference on Acoustics, Speech, and Signal Processing}, 2020,
  pp. 206--210.

\bibitem{mel-gan-kumar2019melgan}
K.~Kumar, R.~Kumar, T.~de~Boissiere, L.~Gestin, W.~Z. Teoh, J.~Sotelo,
  A.~de~Br{\'e}bisson, Y.~Bengio, and A.~Courville, ``{MelGAN}: Generative
  adversarial networks for conditional waveform synthesis,''
  \emph{arXiv:1910.06711}, 2019.

\bibitem{liu2021conditional}
X.~Liu, T.~Iqbal, J.~Zhao, Q.~Huang, M.~D. Plumbley, and W.~Wang, ``Conditional
  sound generation using neural discrete time-frequency representation
  learning,'' in \emph{IEEE International Workshop on Machine Learning for
  Signal Processing}, 2021, pp. 1--6.

\bibitem{kong2021decoupling}
Q.~Kong, Y.~Cao, H.~Liu, K.~Choi, and Y.~Wang, ``Decoupling magnitude and phase
  estimation with deep resunet for music source separation.'' in \emph{The
  International Society for Music Information Retrieval}, 2021.

\bibitem{liu2021cws}
H.~Liu, Q.~Kong, and J.~Liu, ``{CWS-PResUNet}: Music source separation with
  channel-wise subband phase-aware resunet,'' \emph{arXiv:2112.04685}, 2021.

\bibitem{tian2020tfgan}
Q.~Tian, Y.~Chen, Z.~Zhang, H.~Lu, L.~Chen, L.~Xie, and S.~Liu, ``{TFGAN}: Time
  and frequency domain based generative adversarial network for high-fidelity
  speech synthesis,'' \emph{arXiv:2011.12206}, 2020.

\bibitem{vctk-yamagishi2019cstr}
J.~Yamagishi, C.~Veaux, K.~MacDonald \emph{et~al.}, ``{CSTR VCTK corpus}:
  English multi-speaker corpus for cstr voice cloning toolkit,'' 2019.

\bibitem{nu-wave-lee2021nu}
J.~Lee and S.~Han, ``{NU-wave}: A diffusion probabilistic model for neural
  audio upsampling,'' \emph{arXiv:2104.02321}, 2021.

\bibitem{vctk-demand-valentini2017noisy}
C.~Valentini-Botinhao \emph{et~al.}, ``Noisy speech database for training
  speech enhancement algorithms and {TTS} models,'' 2017.

\bibitem{dcase-dataset-mesaros2018multi}
A.~Mesaros, T.~Heittola, and T.~Virtanen, ``A multi-device dataset for urban
  acoustic scene classification,'' \emph{arXiv:1807.09840}, 2018.

\bibitem{choi2020phase}
H.-S. Choi, H.~Heo, J.~H. Lee, and K.~Lee, ``Phase-aware single-stage speech
  denoising and dereverberation with {U-Net},'' \emph{arXiv:2006.00687}, 2020.

\bibitem{lsd-erell1990estimation}
A.~Erell and M.~Weintraub, ``Estimation using log-spectral-distance criterion
  for noise-robust speech recognition,'' in \emph{Proceedings of the IEEE
  Conference on Acoustics, Speech, and Signal Processing}, 1990, pp. 853--856.

\bibitem{pesq-rix2001perceptual}
A.~W. Rix, J.~G. Beerends, M.~P. Hollier, and A.~P. Hekstra, ``Perceptual
  evaluation of speech quality ({PESQ}) - a new method for speech quality
  assessment of telephone networks and codecs,'' in \emph{Proceedings of the
  IEEE Conference on Acoustics, Speech, and Signal Processing}, 2001, pp.
  749--752.

\bibitem{ssim-wang2004image}
Z.~Wang, A.~C. Bovik, H.~R. Sheikh, and E.~P. Simoncelli, ``Image quality
  assessment: from error visibility to structural similarity,'' \emph{IEEE
  transactions on Image Processing}, vol.~13, no.~4, pp. 600--612, 2004.

\bibitem{nu-gan-kumar2020nu}
R.~Kumar, K.~Kumar, V.~Anand, Y.~Bengio, and A.~Courville, ``{NU-GAN}: High
  resolution neural upsampling with {GAN},'' \emph{arXiv:2010.11362}, 2020.

\bibitem{segan-pascual2017segan}
S.~Pascual, A.~Bonafonte, and J.~Serra, ``{SEGAN}: Speech enhancement
  generative adversarial network,'' in \emph{INTERSPEECH}, 2017.

\bibitem{waveunet-macartney2018improved}
C.~Macartney and T.~Weyde, ``Improved speech enhancement with the
  {Wave-U-Net},'' \emph{arXiv:1811.11307}, 2018.

\bibitem{enhancement-qiuqiang-weakly-labeled-kong2021speech}
Q.~Kong, H.~Liu, X.~Du, L.~Chen, R.~Xia, and Y.~Wang, ``Speech enhancement with
  weakly labelled data from {AudioSet},'' in \emph{INTERSPEECH}, 2021.

\bibitem{sspade-zavivska2019proper}
P.~Z{\'a}vi{\v{s}}ka, P.~Rajmic, O.~Mokr{\`y}, and Z.~Pr{\u{u}}{\v{s}}a, ``A
  proper version of synthesis-based sparse audio declipper,'' in
  \emph{Proceedings of the IEEE Conference on Acoustics, Speech, and Signal
  Processing}, 2019, pp. 591--595.

\bibitem{taal2011algorithm}
C.~H. Taal, R.~C. Hendriks, R.~Heusdens, and J.~Jensen, ``An algorithm for
  intelligibility prediction of time-frequency weighted noisy speech,''
  \emph{IEEE/ACM Transactions on Audio, Speech, and Language Processing},
  vol.~19, no.~7, pp. 2125--2136, 2011.

\end{thebibliography}
